# The kinetics of natural ageing in Al-Mg-Si alloys studied by positron annihilation lifetime spectroscopy

J. Banhart[1,2], M.D.H. Lay[3], C.S.T. Chang[1], A.J. Hill[3]

[1] Helmholtz Centre Berlin for Materials and Energy, Hahn-Meitner-Platz, 14109 Berlin, Germany

[2] Technische Universität Berlin, Materials Science and Technology, Hardenbergstr. 36, 10623 Berlin, Germany

[2] CSIRO Materials Science and Engineering, Clayton, Victoria 3169, Australia

The process of natural ageing in pure ternary Al-Mg-Si alloys was studied by positron annihilation lifetime spectroscopy in real time in order to clarify the sequence and kinetics of clustering and precipitation. It was found that natural ageing take place in at least 5 stages in these alloys, four of which were directly observed. This is interpreted as the result of complex interactions between vacancies and solute atoms or clusters. One of the early stages of positron lifetime evolution coincides with a clustering process observed by differential scanning calorimetry (DSC) and involves the formation of a positron trap with around 0.200 ns lifetime. In later stages, a positron trap with a higher lifetime develops in coincidence with the DSC signal of a second clustering reaction. Mg governs both the kinetics and the lifetime change in this stage. Within the first 10 minutes after quenching, a period of nearly constant positron lifetime was found for those Mg-rich alloys that later show an insufficient hardness response to artificial ageing, the so-called 'negative effect'. The various processes observed could be described by two effective activation energies that were found by varying the ageing temperature from 10°C to 37°C.

## I. INTRODUCTION

Al-Mg-Si alloys form the basis of the 6000 alloy series and are the commercially most frequently used Al-based alloys since they can be age-hardened to medium strength while also having many other favourable properties: they can be easily formed, welded, anodised or painted. After solutionising and quenching, these alloys are 'artificially' aged (AA) at typically 180°C. This ageing process has been studied extensively.[1] Ageing at 'room temperature' – the so-called 'natural' ageing (NA) – has received less attention since the strengthening effect during NA is much smaller. The importance of NA lies in the often disturbing negative interactions with a subsequent AA step. This 'negative effect' implies that after an incubation time at 'room temperature' subsequent AA is more sluggish and the peak hardness achieved can be compromised. This effect has been known for a long time[2,3] and has been investigated experimentally or by modelling,[4] but the exact dependence of the 'negative effect' on NA temperature and time and on the alloy composition is not yet fully understood. This is one motivation to study NA. Interestingly, a positive effect can also be found for Al-Mg-Si alloys for Mg and Si contents below 1 wt.% .[5,6]

Natural ageing, especially in Cu-free 6000 alloys, is difficult to study since many methods that have been successfully used for other alloys fail due to low solute content, weak scattering contrast and other limitations[7]. However, the measurement of electrical resistivity, hardness, thermal analysis and positron annihilation lifetime can map the small changes during NA with sufficient accuracy and does allow for studying the kinetics of NA in a phenomenological way. One important result is that NA takes place in distinct stages, see for example Ref. 7 and references therein. Direct observations of the structures formed during NA in the transmission electron microscope (TEM) claimed in the literature[8] could not be reproduced later,[9] and just atom probe (AP) analysis seems to have the potential to reveal the clusters formed. The existence of these clusters is mostly derived by statistical analysis of image data,[9,10,11] while a clear direct visualisation has been reported only for long ageing times and in alloys with high content of alloying elements.[12]

The phenomena observed in Al-Mg-Si alloys after solutionising and quenching at constant 'room temperature' are usually described as two- or three-stage processes comprising various stages in which a given property changes. Clustering of either Si atoms, Mg atoms or both is postulated.[13,14,15,16] Silicon clustering and in later stages its ordering into regular structures[17] is given special importance.

In the present study, we use positron annihilation lifetime spectroscopy (PALS) to detect the subtle structural changes during NA of a series of Al-Mg-Si alloys with varying Mg and Si content. We aim at establishing different regimes of NA by detecting changes of positron lifetimes, to characterise the kinetics of the various reactions at different temperatures and to clarify the influence of Mg and Si content on clustering. PALS is sensitive especially to solute clustering and vacancy-related processes that take place in a supersaturated alloy. We use the PALS technique in such a way that we can monitor changes with a time resolution of ≈2 min and carry out real-time studies. We chose pure ternary alloys based on very pure elements to avoid any interference by further solutes, e.g. Cu that is known to accelerate NA significantly.[18] Moreover the pure ternary alloys are more sensitive to the negative effect than the Cu-containing ones,[3] which will facilitate deriving explanations for this effect in future research. The 'negative effect' also varies with composition in a very sensitive but unknown way[19] which is also a strong argument for using pure ternary systems.

## II. EXPERIMENTS AND PRELIMINARY TESTS

### A. Materials

Pure ternary alloys were prepared from high-purity Al (5N), Mg (4N) and Si (5N) starting materials. Eight different compositions were included in the study, see FIG. 1. The rationale for this choice was to have various alloys with a constant total amount of solute, namely $x_{Mg}+x_{Si} \approx 1.45$ at% (encircled group '1' of alloys 'E','F','G','I') and some alloys with an approximately constant ratio $x_{Mg}/x_{Si}=1$ (group '2' of alloys 'K', 'H', 'L','F'). Another sample very low in Mg was obtained by accidental Mg losses during manufacture (alloy 'J'). Of the alloys used, H is closest to commercial 6060 alloy ($x_{Mg}$=0.36–0.6; $x_{Si}$=0.3–0.6; $x_{Cu}$<0.1), E to 6016 ($x_{Mg}$=0.25–0.6; $x_{Si}$=1–1.5; $x_{Cu}$<0.2), F to 6082 ($x_{Mg}$=0.6–1.2; $x_{Si}$=0.7–1.3; $x_{Cu}$=<0.1), G to 6063 ($x_{Mg}$=0.45–0.9; $x_{Si}$=0.2–0.6; $x_{Cu}$<0.1), all in wt.%.

The alloys were either made in our materials laboratory (code 'L') by melting ≈150 g of the pure constituent metals in a crucible under Ar atmosphere, stirring, water quenching and homogenising them at 550°C for 12 h, or were provided by Hydro Aluminium Bonn (code 'H'), where billets of 10 kg mass were cast, after which the material was homogenised, extruded to 3 mm and cold rolled to 1 mm thickness. Yet another manufacturing route included re-melting alloys made by Hydro Aluminium and drawing single crystals in a crucible (code 'X'). The samples for lifetime measurement were all ≈1 mm thick, and 10 × 10 mm$^2$ in area. The alloy designations used in this paper are composed of the composition code given in FIG. 1 + the manufacturer's code + an experiment number. For example, 'FH16' means alloy 'F', made by Hydro Aluminium ('H') and PALS experiment no. 16.

A microstructure of alloy F after solutionising and quenching is shown in FIG. 2. It exhibits very large crystallites ranging up to 0.5 mm in size. This excessive grain growth is typical for the very pure alloy used.

### B. POSITRON LIFETIME EXPERIMENTS

#### *1. Setup*

The positron lifetime measurements were carried out with two commercial 'Ortec' fast-fast coincidence systems using fast plastic (Bicron) scintillators. In the first spectrometer, the detectors were arranged vertically which facilitates quick mounting of the sample. It was operated at 'room temperature', normally 18±0.5°C. The second spectrometer, arranged horizontally, was equipped with a thin copper sample holder for cooling and heating and was contained in a dry nitrogen atmosphere at all times. The resolution functions (FWHM) of the two spectrometers were 0.250 ns and 0.255 ns, respectively. The channel width of the multichannel analyser was set to 0.05 ns for good statistics, except for a few selected experiments used for tests for possible two-component decays, where this value was changed to 0.0125 ns. A positron source $^{22}$NaCl with initially ≈50±10 µCi (≈1.8 MBq) activity was used. It was placed between a ≈3 µm thick Ti foil folded around the salt without any additional sealing.

## 2. Measurement procedure

For each measurement, a pair of alloy samples was solutionised in an alumina crucible within an air furnace at 535±5°C. After 30 min, the samples were quenched into ice-containing water ensuring that premature cooling was minimised. The samples were immediately dried, stacked in a sandwich with the positron source in between and placed between the two detectors of the spectrometer. Two types of experiments were carried out:

*'In-situ'* measurements: the measurement was started immediately after mounting the sample and spectra were recorded in short intervals, see paragraph II.B.5. In the 'room temperature' spectrometer, the temperature was kept to 18±0.5°C except for one experiment where the 'room temperature' was 27±1°C. The delay between quenching and the beginning of data acquisition ranged between 80 and 160 s. In the spectrometer with the heatable/coolable sample holder, experiments at 10°C, 14°C, 21°C, 35°C and 37°C were carried out. The delay between quenching and data acquisition was longer in this case – about 180 s – since the sample had to be wrapped in Al foil and transferred into the sample holder. Thermalisation of the sample to the measurement temperature was fast (≈30 s) as verified on dummy specimens.

*'Ex-situ'* measurements: here the sample holder of the spectrometer was cooled to −50°C prior to the measurement. After assembling a sandwich containing the 2 samples and the positron source, it was first shock-cooled by placing it in between two Al blocks in contact with a liquid nitrogen bath and then mounted. Data was acquired for an extended period to ensure good statistics. After this, the sample was removed from the sample holder and pressed between two Al blocks serving as 'room temperature' heat reservoir. After a given time at 18°C, the measurement procedure was repeated. The rationale of this experiment was to preserve a given state of NA and to acquire more data than possible in the 'in-situ' experiments.

## 3. Data analysis

The program 'LT9' was used for analysing the PALS data.[20] The essentials of analysis are modelling the resolution function of the system either by a single Gaussian or by a Gaussian and two exponential tails. Moreover, the source corrections were subtracted, see below. In all cases, a good fit could be obtained by using a single exponential for the sample which represented an average lifetime $\tau_{av}$. Use of a single Gaussian for the resolution function led to more stable results and almost the same values for $\tau_{av}$ as more elaborate spectrometer resolution models and was therefore preferred.

## 4. Source corrections

As part of the positron annihilation takes place in the source itself (NaCl, Ti foil), a contribution has to be subtracted that represents those contributions, see Ref. 21. Source corrections were determined by measuring well annealed samples of 99.999% pure Al. In addition, other annealed pure elements (Mg, Si, Cu, Ni, In) of various purities were characterised. Three lifetime components were assumed: the expected bulk lifetime for the element considered, a lifetime component around 0.4 ns representing the source material and a long lifetime component around 2 ns representing the influence of interfaces and surfaces. We tried to derive a consistent set of source corrections from all the different elements, but found that it is hardly possible to find values that satisfy all boundary conditions perfectly. We then decided to base the corrections on experiments on the pure aluminium sample only. It is found that the spectra for such specimens could be expressed by the three components mentioned and that in some cases even fits with 5 free parameters (3 lifetimes + 2 intensities) could be performed. We used two different sources for the work presented here, and derived two corresponding sets of source corrections:

- sample group 1: 5% of a 0.380 ns component + variable percentage of a 2 ns component, usually 0.1 – 0.3% (full symbols in FIGS. 4, 6–11),

- sample group 2: same as sample 1, but 4% of a 0.400 ns component (open symbols).

We did not fix the intensity of the 2 ns component, but allowed it to fluctuate around a mean value. Using precise source corrections is a prerequisite to measure correct absolute values and to be able to decompose lifetime spectra into various components. However, they have little influence on the relative changes observed during ageing as is illustrated in FIG. 3, where different source corrections were applied to the same data set. Obviously, the overall course of the average lifetime is the same for all lifetime corrections (discussion will be given later). The variances of the different fits show a clear improvement when the 2 ns component is added and a further improvement when the component related to annihilation in the NaCl is taken into account.

### *5. Feasibility of fast measurements*

As the objective of the present work was to measure positron lifetimes with a high time resolution, experiments were carried out to determine the minimum suitable accumulation time that allows for a reliable determination of the kinetics of NA. For this, following quenching of an alloy test specimen, we acquired data every 45 seconds sequentially. By binning either 3 or 9 data sets into one new set we simulate longer accumulation times. FIG. 4 shows that for 45 s of acquisition (5000 peak counts, ≈50000 total counts) large statistical errors are found but an overall trend (discussed later) is clearly visible. Binning 3 or 9 spectra into one new set both reduces the statistical errors and confirms the trend given by the unbinned data. We decided to go for a compromise and to use about 2 min of acquisition (15000 peak counts) in our study and trust that we can resolve the main features of average lifetime change very well.

### *6. Presence of distinct lifetime components*

For Al alloys that contain a high defect density after quenching one mostly assumes complete trapping into defects with no or only a small bulk component, e.g. Refs. 22,23. As there has been a report on the occurrence of two lifetime components immediately after quenching[24] we tried to decompose some of our spectra into two lifetime components, one related to bulk annihilation (ansatz: ≈0.160 ns), one to the average of all the defects (ansatz: ≈0.230 ns). It was found that the fit of two decaying exponentials to various data sets did not reduce the variance significantly and also led to instable fits, which is why we consider the presence of two lifetime components unlikely, see FIG. 5.

## III  RESULTS

### A  In-situ positron lifetime measurements during NA at 18°C

#### *1. Alloy F*

Average positron lifetimes for alloy F were measured in-situ during NA for 8 different specimens as given in FIG. 6(a). In addition, lifetimes measured ex-situ at –50°C have been added. As the in-situ measurements involved acquiring data in constant time intervals, there is a strong accumulation of points towards longer times on the logarithmic time scale, which obscures the trend, see FIG. 3. Therefore, the measured data was averaged such that an approximately equidistant mesh on the logarithmic scale was obtained as demonstrated in FIG. 3, lower curve. This was also done for the data displayed in FIGS. 7, 8, 10 and 11.

The first lifetime measured 2 to 3 minutes after quenching typically ranges from 0.221 to 0.230 ns, with a pronounced scatter between the individual experiments. There is a short initial *stage I* in which the lifetime is constant or only slowly decreases as can be seen best from FIG. 6(b) that contains early data only and uses a linear time scale. Of all the measurements shown, only one sample (FH11) drops continuously, whereas the other data show a change of slope after about 5 to 10 min. After this, the average lifetime decreases again and reaches minimal values between 0.212 and 0.215 ns after about 50 min. This decrease we call *stage II*. At the end of this stage all the individual measurements have merged into one curve, i.e. reproducibility is very good. This applies also to the alloys FH and FL that were manufactured in a different way but have a similar composition. By fitting 3$^{rd}$ order polynominals to $\tau_{av}(\log(t))$ in the region of the minimum, we obtain an averaged 50±7 min for the end of stage II. After passing the minimum, an increase of average lifetime is observed up to 760±60 min after quenching (obtained by polynominal fitting too) where a maximum value of 0.217 ns is reached. This increase we call *stage III*. In the final *stage IV* the lifetime drops again. We could not follow this stage for all the alloys but found an individual value of 0.214 ns after $8 \times 10^4$ min of NA. The as-received sample before any heat treatment exhibited a positron lifetime of 0.213 s, which suggests that this is also the asymptotic value for the positron lifetime after very long NA.

## *2. Alloy H*

The positron lifetime evolution for alloy H is shown in FIG. 7(a). It looks very similar to that of alloy F. One difference is that the minimum after stage II and the maximum after stage III occur later, namely 84±12 and 900±60 min after quenching, respectively. Another difference concerns stage I. In contrast to alloy F, no stage of approximately constant lifetime can be observed. On the contrary, all the 4 measurements displayed in FIG. 7(b) on a linear time scale show a negative slope from the first to the second point that is 3 times larger than the average over the first 20 min, i.e. the lifetime change slows down after 4 – 6 min.

### *3. Other alloys*

Alloy G has already been presented in FIG. 4 in the context of test measurements. Data for further alloys (E, J, K, L, I) were measured as well. A representative curve for each alloy is included in FIG. 8. The various curves can be grouped as there are common features.

A: alloy J is basically alloy H from which most of the Mg has been removed during single crystal preparation. The positron lifetime decreases monotonically after quenching for more than 3000 min. It levels off after as seen by a value measured after $4.3\times10^4$ min (1 month). The data can be fitted to a straight line very well on the logarithmic time scale.

B: alloy K shows a lower initial positron lifetime that decreases almost linearly on the logarithmic time scale. We observe a minimum after 165±15 min and an increase to a maximum that is reached after 2670±250 min, followed by a slow re-decrease.

C: alloys H, E and L show an even lower initial lifetime value. E and L behave similar to the alloy H discussed in the previous section. The lifetime decrease levels off after the first 2 – 3 points. A minimum is reached after 84±12, 86±5 and 76±5 min (for H, E, L), a maximum after 900±60, 927±25 min (for H, L). The polycrystalline and single crystalline samples do not show a notable difference. The kinetics of positron lifetime evolution for all these alloys is similar.

D: alloys F, G and I show a further reduced initial lifetime. A regime of initially constant lifetime (stage I) is found for alloys F and G. It is even more pronounced for alloy G, see also FIG. 4, while for alloy I its existence cannot be decided. The minimum of lifetime is reached after 50±7, 51±4 and 64±4 min for F, G and I, respectively.

**B. Ex-situ lifetime measurements at −50°C**

FIG. 9 shows the positron lifetime evolution at –50°C for an alloy H sample. Between the individual data sets marked by vertical bars, the sample was aged at 18°C for the time specified. Obviously, the positron lifetime is constant, indicating the absence of significant microstructural changes at –50°C. The average of the data for each segment has been added to FIG. 7 together with the results of another series of this kind. The same was done for alloy F, see FIG. 6. It is clear from these figures that the lifetimes measured ex-situ agree very well with the in-situ measured values.

## C. In-situ positron lifetime measurements during ageing at different temperatures

FIG. 10 shows the evolution of positron lifetime for alloy H at different temperatures directly after solutionising and quenching. The measurement at 18°C is reproduced from FIG. 7. We observe that the lifetime curves evolve in a similar way for all the temperatures in the applied range from 10°C to 37°C. For 37°C, the evolution is too fast to capture the initial decrease, but only at this temperature the curve gets flat for longer times, thus indicating that an asymptotic value is reached. For 18°C and 26°C, the data acquisition rate was higher since the detectors were closer together in the spectrometer without the temperature-controlled stage and therefore the density of points is higher. Alloy F (not shown) exhibit a trend very similar to alloy H.

# IV DISCUSSION

## A. Comparison with the literature

Only few experimental PALS studies of NA in 6000 alloys have been published. Recent measurements[25] are compared to the current ones in FIG. 11. In those measurements, an initial increase of lifetime is observed, starting from a first point measured after 15 min. This increase corresponds to stage III of the present work. For longer times, there is a slight indication for a stage IV. Unfortunately, around the suspected maximum of lifetime (see broken lines) no data is given. What is definitively not seen in their measurements is the initial decrease (stage II), although their data collection time (initially ≈15 min) should have allowed them to detect it. As their error bars are large, the lifetime decrease could be hidden in the noise. The alloy composition was only slightly different from our alloys E and F, for which corresponding lifetime curves are given in FIG. 11 for matters of comparison.

Egger et al.[22] investigated 6082 alloys naturally aged for >6 days with a pulsed positron beam and find complete trapping into defects with a lifetime of 0.223 ns, a value which is 0.007 ns higher than our corresponding value for the similar alloy F.

Klobes et al.[24] measured positron lifetimes for our alloys FL and HL ex-situ, as described in Sec. 3.2. They observe an initial increase from very low values (0.145 and 0.180 ns for FL and HL, resp.) up to 0.220 to 0.230 ns after just 10 minutes, after which the lifetime remains constant. Moreover, two lifetime components are detected within the first 5 to 10 min. These results are in clear contradiction to this work and we do not have an explanation for this.

Buha at al.[26] measured the NA behaviour for alloy 6061. They found a rapid decrease from initially 0.231 ns to 0.221 ns within about 100 min which looks similar to stage II found in this work. However, neither stage III nor stage IV are clearly visible. A possible reason is the presence of 0.25% Cu in the alloy investigated which might modify the ageing behaviour. The absolute values are higher, perhaps because no source correction was subtracted.

In summary, some of the features measured in this work can be found by inspecting literature data, but none of the references reports all four stages. It is worth noting that in Al-Mg-Zn alloys a non-monotonic evolution of lifetime has been reported recently,[27] unlike earlier work on other alloys where the decrease of lifetime was always monotonic under isothermal conditions.[23]

**B. Interpretation of different stages of positron lifetime evolution**

*1. General considerations*

The observed complex pattern of positron lifetime changes during NA is caused by the continuous evolution of both the vacancy and the solute configuration in the supersaturated alloy during which the number of annihilation sites and the electron density around those sites change. The measured global lifetime expresses an average over a spectrum of different local lifetimes each of which corresponds to a specific site. Due to the experimental restrictions – too low a time resolution due to intrinsic restrictions of the detector system and too low a count rate due to the real-time character of the experiment – a deconvolution of the average lifetime into individual lifetimes is impossible. This problem is common in such studies, see e.g. Ref. 23. The decay spectra corresponding to many not too different positron lifetimes are smeared out to one single decay. To further complicate the situation, only the positron lifetimes due to annihilation in the simplest positron traps are known theoretically. Therefore, discussions of measured lifetime evolutions found in the literature and also in the present paper are based on assumptions regarding the positron lifetime of possible positron traps in the material.

For pure Al, the situation is well investigated. In annealed alloys with few point defects, most positrons annihilate in the bulk matrix and have a lifetime between 0.160 and 0.170 ns.[28] If positrons annihilate predominantly at vacant lattice sites, the lifetime is known to be around 0.250 ns.[28] Immediately after quenching pure Al metal from a high temperature, e.g. 500°C, there are so many such vacant sites that virtually all positrons are trapped there and the average lifetime is around 0.250 ns. During NA, the measured lifetime drops and approaches values close to the bulk lifetime for prolonged ageing since vacancies go to sinks after diffusing through the pure bulk and the fraction of positrons annihilating in the bulk increases. We have verified this by performing two-component fits.

In our alloys the situation is different. After quenching, the vacancies start diffusing through the Al bulk. As the concentration of solute atoms is high – 0.8 to 1.5% in our case, which is two orders of magnitude higher than the vacancy concentration –, a vacancy encounters a solute atom already after some 100 site changes. The jump rate $\Gamma$ of a vacancy in a pure Al matrix is given by:[29]

$$\Gamma = Z\nu e^{-H_m/RT}, \qquad (1)$$

where $Z = 12$ is the coordination number of the vacancy in the fcc lattice, $\nu$ is the atomic jump frequency and $H_m$ is the migration enthalpy of an Al atom. Reasonable values for $H_m$ range around 64 kJ/mol, as estimated from experimental values for the activation energy of self diffusion in Al (120 – 142 kJ/mol)[30] and the enthalpy of vacancy formation (60 – 74 kJ/mol),[31] the latter agreeing well with calculations.[32] In Eq. 1, we use a theoretical value $\nu \approx 2\times10^{13}$ s$^{-1}$ (Ref. 32) and obtain a jump frequency of $\Gamma \approx 800$ s$^{-1}$ at T=18°C. Even when using slightly lower or higher values for $H_m$, the result is that long before we have measured the first positron lifetime, most vacancies have already been in contact with solute atoms and, depending on their binding energy with those, they may remain with the solute for some time. Solute diffusion sets in and solute agglomeration may occur. It has been concluded indirectly, that vacancies must be able to detach from solute atoms and clusters to transport other solute atoms since the vacancy density is just 1/100 of the solute density. This is called the *vacancy pump* model.[33] However, it has been argued that vacancies spend most of the time attached to solutes[34] so that free vacancies would play little role for positron trapping in this stage.

### 2. Initial (unobserved) positron lifetime decrease (stage 0)

Immediately after quenching and not observable with our lifetime setup, positrons should annihilate in the still free vacancies, i.e. the lifetime should be around 0.250 ns. It takes little time ($\leq$ 2 min) to knock down this lifetime to 0.215–0.235 ns for most of the alloys. In the low-solute alloy H, the tail of this fast initial decrease that slows down after a few minutes might be observable, c.f. FIG. 7(b), but only in alloy J most of the lifetime decrease, starting here from 0.247 ns, can be observed. In the alloys that contain more solute – alloys F, G and I – the initial decrease is too fast to be resolved. In the alloy series K→H→L→F, both the Mg and the Si content increase, in the series E→F→G→I, Si is replaced by Mg in steps of 0.2% while keeping the total amount of solute constant. In both series, there is a tendency for an increasingly lower first measured lifetime (2 to 3 min after quenching), indicating that the higher chance of vacancies to encounter solute atoms is important but also that Mg controls the kinetics of the initial positron lifetime decrease.

*3. Possible mechanisms for the observed positron lifetime decrease in stages 0 and II*

A prominent feature of the experimental lifetime curves in FIGS. 3, 4, 6–8 and 10–11 is the initial decrease which implies a problem since single-component positron lifetime spectra with $\tau_{av}$ as low as 0.200 ns are difficult to explain. In Al-Mg-Cu alloys the high electron density caused by Cu-attachments to vacancies are held responsible for short lifetimes.[23] In Al-Mg-Si alloys, such an effect is expected to be less efficient. Preliminary first-principles calculations have confirmed that the effect of Mg or Si decorations of vacancies is small and that one Si atom decreases positron lifetime, while Mg increases lifetime, but both values are within only a few 0.001 ns of the value for vacancies in a Al matrix, 0.250 ns. The exact numbers depend on the way lattice relaxations are treated.[35] This means that only Si can decrease the lifetime of a positron trapped by a vacancy. If more than one Si or Mg atom agglomerated end eventually formed clusters around a vacancy, this trend could get even stronger, depending on how strong lattice relaxation, electronic structure changes or changes of the local crystal structure, e.g. to that of a GP zone, are, but such calculations are not available for Al-Mg-Si. TABLE 1 summarises the positron lifetimes associated to such traps.

In Al-Cu alloys, one Cu atom attached to a vacancy was calculated to lower the positron lifetime by 0.003 ns, while eight Cu atoms reduce the lifetime by 0.016 ns.[36] Another calculation finds that 4 Cu atoms attached to a vacancy even reduce the lifetime by 0.019 ns.[37] Cu decoration of vacancies therefore has a stronger effect on positron lifetime than Si decoration which is why the lifetimes in Al-Cu alloys are generally shorter after NA,[23] but even here, the theoretical calculations underestimate the observed positron lifetime change reported, e.g. in Ref. 23.

Further potential positron traps are clusters or coherent precipitates that neither contain vacancies nor are attached to such. Mg and Si atoms have much stronger positron affinities than Al and therefore it can been argued that precipitates enriched in Si and/or Mg in an Al matrix may trap positrons if they exceed a given volume which can correspond to only few atoms for high Mg and Si contents.[38] The lifetime of positrons in such traps has been described by averages of the individual elements.[39] This simple approximation leads to a positron lifetime of 0.200 ns for a cluster containing, e.g., 60% of Mg and Si in equal parts and 40% Al, see TABLE 1. Small clusters should be shallow traps just a few tens of meV deep, unlike vacancy-related potentials that are 1.9 eV deep in Al.[40] Positron lifetimes in shallow traps should be temperature dependent. Such a dependence has recently been shown by experiments at temperatures varying from −233 °C to −40°C on alloys HL and FL that had been naturally aged for different times.[41] Our experiments at 18°C and −50°C yielded similar results, see FIG. 6 and FIG. 7, which allows to conclude that there is little temperature dependence above −40°C. The observed temperature dependence levelled off after a few hours of NA, indicating that either the concentration of the shallow traps had decreased or they had become deep traps.

Recent atom probe experiments on Al-Mg-Si-Cu alloys have shown that after 210 min of NA a mixture of small (< 7 atoms), intermediate (up to 166 atoms) and very few larger clusters exists.[42] After shorter NA the clusters would be smaller but still could have a sufficient positron affinity to trap positrons at room temperature due to the high affinity of both Mg and Si.[38] Thus, the model of a population of vacancy-free clusters growing during NA provides a second explanation for the decrease of positron lifetime in the course of NA. As the solute concentration is 100 times the vacancy concentration the number density of such clusters could easily be in the range of the vacancy density even if each cluster contained tens of solute atoms.

The calculated values for Al-Cu alloys[36,37] suggest that exclusive annihilation of positrons in vacancies decorated by Mg or Cu atoms cannot explain the lifetimes described in Ref. 23, but that the lifetimes in the traps postulated there (0.203 and 0.213 ns) were also influenced by vacancy-free clusters as in the present work.

Theoretically, lifetimes decreasing down to 0.198 ns could also be explained by a bulk annihilation component (0.160 to 0.170 ns) growing in intensity at the cost of the other positron traps. This would happen if many vacancies were absorbed by solutes or sinks and positron trapping would no longer be saturated. However, this should become evident as a second component in the measured lifetime spectra. Neither in this work nor in most of the literature have two-component decay spectra in 6000 or other concentrated alloys been consistently observed.[22,43] Two lifetimes within the first 10 minutes after quenching of alloys F and H have been reported,[24] but this could neither be verified (see Sec. II.B.6) nor would provide explanations since the measured lifetime continues to decreases even after 10 min.

In view of the ambiguity of the situation we adopt the pragmatic viewpoint that NA after solutionising and quenching gives rise to the formation of a mixture of different traps with a spectrum of lifetimes down to below 0.200 ns, which gives rise to single component positron spectra with lifetimes as low as 0.199 and 0.198 ns (see FIG. 8 and Ref. 7, resp.).

In stage II, starting from values between 0.215 and 0.247 ns – depending on the alloy – the lifetime drops to values as low as 0.199 ns. As after stage 0 the lifetime has already dropped to a low value for the Mg-rich alloys, the slope of the lifetime decrease is smaller in stage II for those alloys. However, the final lifetime value reached before it begins to increase again is very similar for all the alloys. This indicates that at that point the development of the positron trap postulated above has come to an end. Thermal analysis of alloy F by DSC has shown that various types of clusters develop during natural ageing.[44] One type of cluster (called C1) is completed after 60 min, while a second type of cluster (called C2) starts forming at some instant between 30 and 75 min after quenching and continues growing for more than a week. As the positron lifetime minimum is reached after 50±7 min for alloy F, it is near at hand to postulate that it is cluster C1 that traps positrons and gives rise to the observed low lifetimes down to ≈0.200 ns. FIG. 8 shows that alloy F is the alloy in which the lifetime minimum is reached first. The balanced Si:Mg ratio of ≈1.25 and the higher level of Mg compared to alloys H and K obviously facilitates the formation of cluster C1.

*4. Lifetime increase – stage III*

All but one alloy show a re-increase of the measured lifetime after the minimum. The experimental data supply a direct proof that this increase is linked to the presence of Mg since, (i) it is not observed for very low Mg content, see FIG. 8, and (ii) a higher content of Mg leads to a faster increase of positron lifetime and makes it reach the maximum value at the end of stage III earlier as shown in FIG. 12. In view of the results obtained by DSC where cluster C2 was shown to develop precisely in this stage, the increase of lifetime can be explained by assuming that cluster acts as a positron trap. Since no dissolution of C1 has been observed,[44] C2 must contribute with a higher positron lifetime than C1.

In the literature, an increase of positron lifetime has also been observed for Al-Mg-Zn alloys – however, without the preceding decrease and already within the first hour after quenching.[27] That increase has been explained by fast aggregation of Mg to vacancies already decorated with Zn atoms which appears analogous to the situation here. The interpretation was also supported by coincidence Doppler broadening (CDB) measurements. As calculations show that a single Mg atom increases the positron lifetime in a vacancy, Mg aggregation to Si-rich clusters around vacancies is likely to have the same effect. Mg aggregation to vacancy-free clusters – though not accessible to calculations – should have the same effect.

## 5. Final lifetime decrease – stage IV

After reaching a maximum, the lifetime starts to decrease again. One explanation is that the nucleation stage of cluster C2 eventually ends and some clusters C2 start growing at the expense of others, i.e. the clusters coarsen. The intensity of the lifetime component associated to C2 would then decrease and that of cluster C1 (lower lifetime) regain importance. Another reason could be zone formation and ordering phenomena occurring within the clusters that could lower the intrinsic positron lifetime. Ordered mono-layered zones, 2.5 nm thick and 30 nm long, have been observed in Al-Mg-Si samples aged at 70°C,[45] and smaller precursor versions of such zones could be formed already at room temperature, invisible in TEM but influencing positron lifetimes. In Al-Mg-Zn alloys, a decrease of positron lifetime has also been observed during NA, however already after 1 hour.[27] The explanations offered there include an further enrichment in Zn and a reduction of positron trapping into defects. It would be too speculative to conclude which mechanism leads to the lifetime decrease in our case.

## 6. Stage I

A short stage of nearly constant lifetime occurs in two alloys containing 0.6 and 0.8 at.% Mg. As stage I is not observed for alloy E and is not clearly seen for alloy I either, a Si:Mg ratio close to 1 promotes this stage, indicating that the interactions between Mg and Si are responsible for it. DSC experiments have revealed a very small and early clustering peak called C0 in Ref. 44 which could be associated to stage I. The length of stage I strongly depends on temperature. In alloy F, it lasts for about 40 min at 10°C, for about 20 min at 14°C, compared to less than 10 min for 18°C, i.e. the process is thermally activated.

There seems to be a connection between the presence of a pronounced stage I and a strong negative effect of NA on subsequent AA. According to Ref. 46, reproduced in Ref. 47, alloy F and G lose more than 20 MPa of strengthening potential when they are naturally aged for 24 hours before AA (15 h at 165°C), alloys E and L lose just 10 MPa, whereas all the other alloys investigated even show a positive effect. In alloy F, the full negative effect is established after about 18 min of NA,[7] suggesting that the processes observed in stage I are a prerequisite for the negative effect.

## 7. Further observations

From the measurements presented, little difference between polycrystalline and single crystalline samples could be derived. As the polycrystalline samples exhibited very large grains due to excessive grain growth during solutionising, it is not surprising that they behave similar to single crystals.

The measurements at different temperatures, see FIG. 10, show that the processes causing the positron lifetime changes are thermally activated as the overall course of the lifetime evolution is similar for the entire temperature range. To some extent, the addition of Mg to an alloy has an effect similar to increasing the temperature, namely shifting the features of the lifetime evolution to shorter times. The presence of Mg therefore promotes diffusion as an increased temperature also does.

## C. Activation energy analysis

From the temperature-dependent measurements, estimates for the activation energy of the various processes can be calculated. Other than the signals provided by thermal analysis, positron lifetimes are not related to the volume fraction of a phase. We therefore cannot correlate the measured lifetime to a fraction of converted phase and, e.g., analyse clustering kinetics with an Avrami-type equation.

We adopt the following viewpoint: accepting the above interpretation of positron lifetimes, the change from stage II to stage III occurs whenever the environment of the positron traps has changed in a defined way, e.g. a certain enrichment in Si has taken place. This point of minimum positron lifetime represents a certain stage of the phase transformation and it can be assumed that it corresponds to an unknown but fixed converted volume fraction $f_{II \to III}$, the value of which would be the same at all temperatures. Applying the Johnson-Mehl-Avrami equation for heterogeneous nucleation and growth one can write for isothermal conditions:[48]

$$f_{II \to III} = 1 - e^{-(\kappa(T) t_{min})^n} = const, \qquad (2)$$

where $t_{min}$ is the time of the lifetime minimum, $\kappa(T)$ the temperature-dependent reaction rate and $n$ the Avrami exponent. Combining Eq. (3) and an Arrhenius ansatz for the reaction rate, $\kappa(T) = \kappa_0 \exp(-Q/RT)$, we obtain:

$$\ln t_{min}^{-1}(T) = \frac{1}{n} \ln\left[\ln\left(\frac{\kappa_0}{1-f}\right)\right] - \frac{Q_1}{RT}. \qquad (3)$$

Therefore, an activation energy $Q_1$ can be determined graphically without knowledge of $f$, $\kappa_0$ and $n$. The same applies to the transformation from stage III to IV. Here, we can either use $t_{max}^{-1}$ or $(t_{max}-t_{min})^{-1}$ in Eq.(4) as measures for the reaction rate, depending on whether one assumes that the reaction leading to the lifetime increase starts directly after quenching in competition to the reaction decreasing lifetime, or whether this reaction follows the first one. We obtain a value $Q_2$. FIG. 13 shows the corresponding data for alloys H and F. The activation energies $Q_1$ and $Q_2$ determined in this way are listed in TABLE 2. As the processes observed are quite complex, these energies might not resemble the activation energies known from simple reaction kinetics but rather represent weighted averages over various such activation energies. The values for $Q_2$ obtained for the maximum of the lifetime curve are very similar for the two variants applied since $t_{min} \ll t_{max}$. For both the alloys $Q_1<Q_2$ holds, i.e. the first reaction is easier to activate by a temperature increase.

The analysis carried out here is similar to what has been applied to Al-Cu-Mg before, where various isothermal positron annihilation experiments were performed at different temperatures, after which the curves were collapsed onto one master curve.[23] Such analysis implicitly assumes that the shape of the positron lifetime curve does not change with temperature, which is equivalent to the assumptions used in this work.

Cluster activation energies for 6000 alloys have been determined by many authors both by isothermal analysis – based on electrical resistivity – or by constant heating rate analysis applying DSC. The scatter of data is very large for both variants. Isothermal analysis yielded 72 kJ/mol for a Al-0.9%Mg-0.51%Si alloy in one case,[49] but only 45–49 kJ/mol for a pure ternary Al-0.46%Mg-1%Si,[25] 39 kJ/mol for a ternary balanced Al-0.5%$Mg_2$Si alloy,[13] and 43 kJ/mol for a 6111 alloy.[50] The same applies to DSC analysis at various constant heating rates, usually based on the Kissinger equation or a variant of this. 6061 alloy: 33 kJ/mol,[51] Al-0.8%Mg-0.9%Si: 44 kJ/mol,[14] Al-0.98%Mg-0.58%Si(Cu,Fe,Cr): 79 kJ/mol,[52] various pure Al-Mg-Si: 52–65 kJ/mol,[53] Al-1%Mg-Si with various Si contents: 31–79 kJ/mol.[54] Obviously, the values given – including ours – cover a wide range although the alloys and methods of analysis are rather similar. Usually one tries to explain such results in terms of the migration enthalpies of either Mg or Si in Al (both $\approx$60–70 kJ/mol) and either claims that a high binding energy between Mg and Si is responsible for lower values found[14] or that vacancy clustering shifts the activation energy. Altogether, such discussions are nor based on a clear-cut definition of an activation energy, which, together with deficiencies of data evaluation, could give rise to such huge discrepancies. Our results agree well only with the highest values given in the literature. A more thorough analysis of different DSC measurements and analyses is required to identify the reasons for the experimental scatter.

## V. CONCLUSIONS

Positron lifetime measurements show that after solutionising and quenching, the subsequent natural ageing process at temperatures between 10°C and 37°C shows the features schematised in FIG. 14:

- **Stage 0**: there is only indirect evidence for a rapid initial decrease of positron lifetime from a pure vacancy-related value around 0.250 ns to the initial values measured after 2 – 3 min. The decrease is caused by the rapid formation of vacancy-solute complexes and the onset of clustering. The process is controlled by the total amount of solute and especially by the Mg content. The few available theoretical predictions suggest that it is rather Si than Mg that neighbours vacancies in such traps.
- **Stage I**: a transient regime of constant lifetime was observed for alloy F and even more clearly for alloy G, but not for the other alloys. The duration of stage I was <10 minutes at 18°C. In this stage, the negative effect on subsequent artificial ageing is largely established.
- **Stage II**: positron lifetime continues to drop, reaching a minimum value after typically 48 – 165 min at 18°C for most of the alloys. A first clustering reaction C1 as detected by DSC takes place and gives rise to the development of positron traps with a lifetime of around 0.200 ns or even below.
- **Stage III**: an increase of positron lifetime is observed until a maximum value is reached after 500 – 1200 minutes at 18°C, depending on the alloy. The clustering reaction C2 detected by DSC can explain this. C2 gives rise to a kind of trap in which positrons have a longer lifetime than when they are trapped by C1. Mg controls the kinetics of clustering and is responsible for the lifetime increase due to aggregation to clusters.
- **Stage IV**: a re-decrease of positron lifetime sets in. Coarsening or ordering of cluster C2 are possible reasons.

The observed positron lifetime changes cannot be explained by annihilation in vacancies alone. Additional annihilation in vacancy-free clusters must take place.

All the processes observed are thermally activated and can be described by two effective activation energies that all range from 74 to 97 kJ/mol.

For the future, positron Doppler broadening (DB) experiments could provide additional independent evidence for the clustering effects observed by PALS. Preliminary DB experiments revealed that the effects expected are very small and that a low-background technique such a coincident DB[43] or high-momentum analysis of single-detector DB data[55] is essential. However, recent calculations have shown that a quantitative analysis of effects of Mg and Si might be out of reach since these elements are neighbours of Al in the periodic table.[36] A more reliable calculation of positron lifetimes and momentum distributions as well as binding energies between vacancies and solute clusters would facilitate the interpretation of experimental PALS data, but such calculations are currently not available. Finally, refined cluster identification procedures by atom probe are currently being developed[56,57] and could provide reliable information on sizes and compositions of clusters forming during NA.

## Acknowledgements

We would like to thank Prof. J. Hirsch of Hydro Aluminium, Bonn for providing some of the alloys, Dr S. Pas, CSIRO Clayton, Dr. M. Haaks, B. Klobes and B. Korff, HIKSP Bonn, and Prof. G. Amarendra from IGCAR Kalpakkam for their advise and encouragement concerning positron annihilation experiments, Dr. K.-D. Liss and K. Yan, ANSTO, Lucas Heights, for providing calibration materials and Dr. G. Behr from IFW Dresden for preparing single crystals.

# Figures

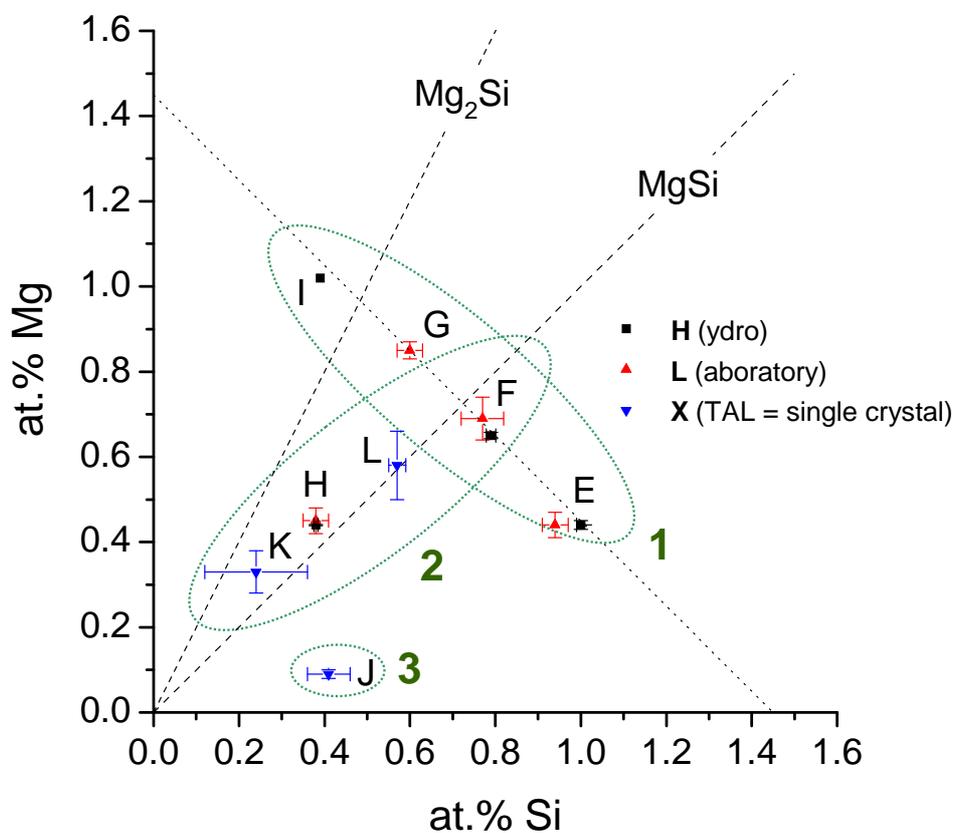

FIG. 1. Ternary alloys used for the present study. Compositions were determined using sparc OEM. Each alloy composition is identified by a capital letter. The different symbols denote source of the material: Hydro Aluminium Company, own laboratory or XTAL (= single crystal) grown from melts based on Hydro alloys and pure 5N aluminium.

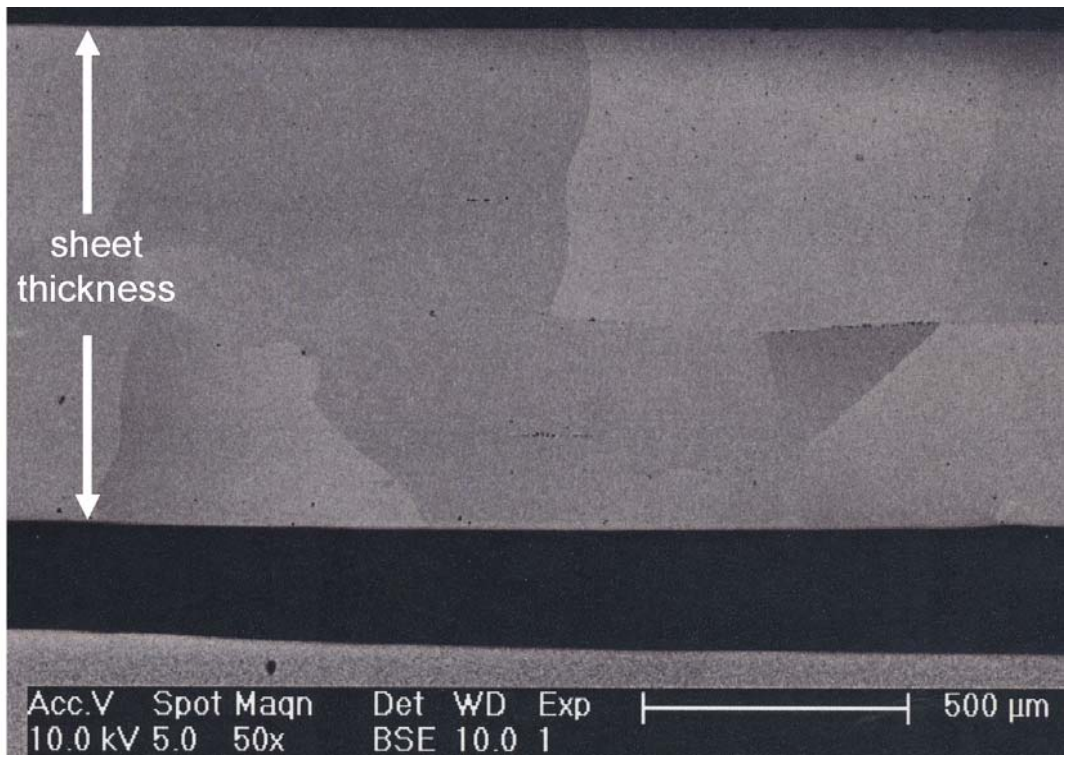

FIG. 2. Microstructure of alloy F after solutionising at 535°C for 30 min and quenching.

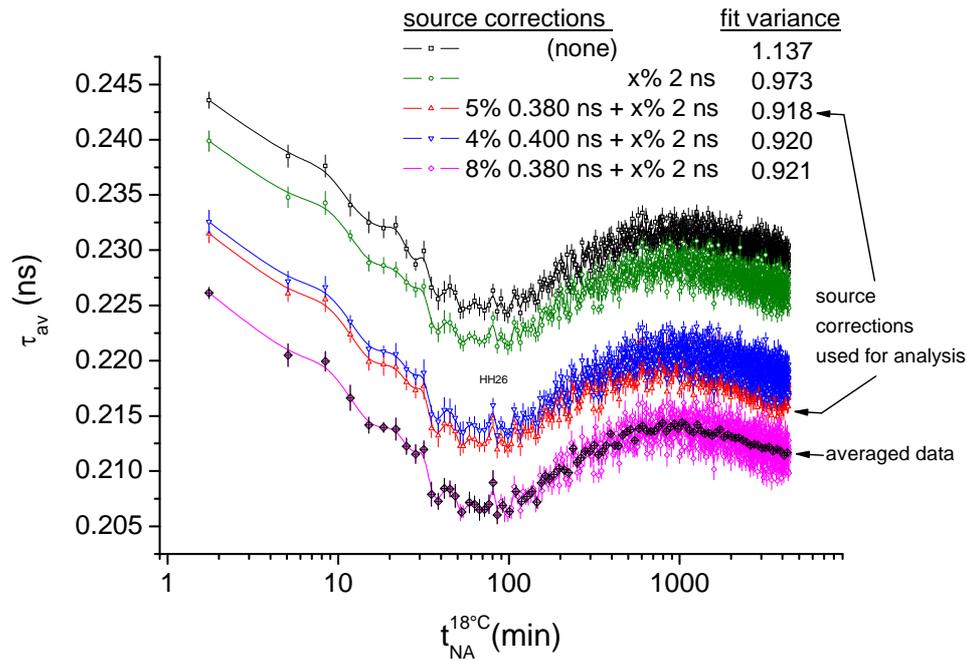

FIG. 3. Application of different source corrections to a data set representing alloy H during NA. The different curves were obtained by subtracting the contribution of up to 2 different annihilation lifetimes. In the legend, 'x% 2 ns' means that the intensity $x$ for this component was not fixed, but it was allowed to fluctuate by ±25% around a mean value. The same applied to the FWHM of the spectrometer resolution function (single Gaussian) which was allowed to fluctuate about ±0.005 ns around the average value of 0.250 ns. For the lower curve (8% 0.380 ns) the averaging procedure is demonstrated. It leads to data on an approximately equidistant mesh on the logarithmic scale (crossed diamonds), superposed on the original data.

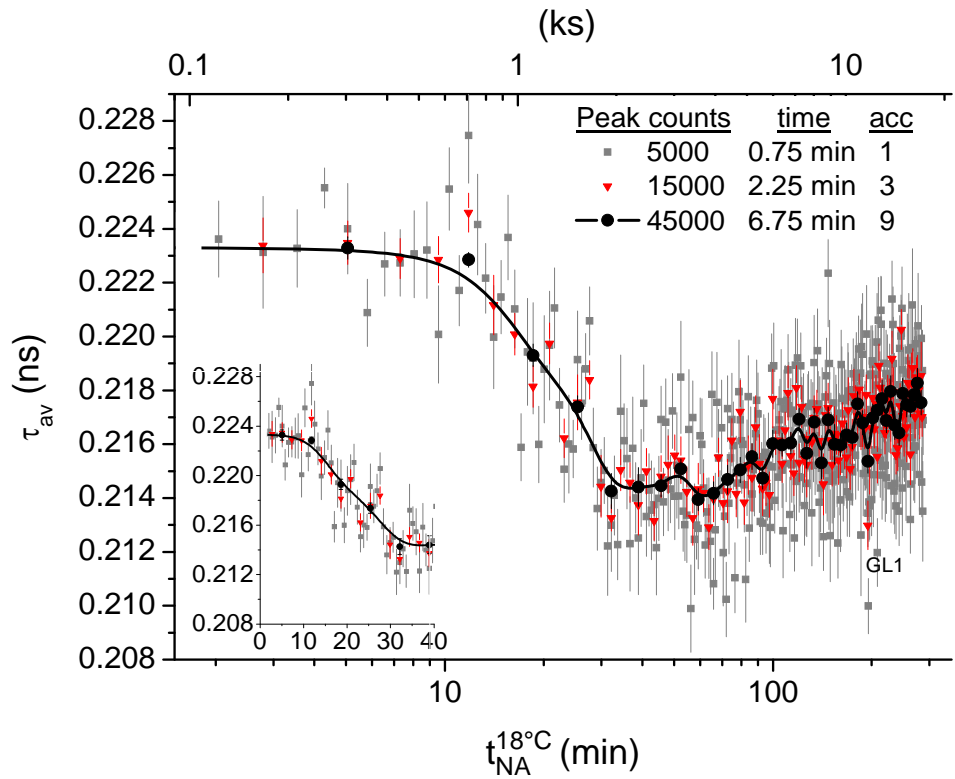

FIG. 4. Evolution of average positron lifetime in alloy G during NA directly after solutionising and quenching. Three analyses are shown: that of the original data where each spectrum was acquired during 0.75 min (45 s) and two analyses where 3 or 9 data sets were accumulated. The solid line is a spline interpolation of the data representing the longest acquisition time. The inset magnifies the first 40 minutes on a linear time scale.

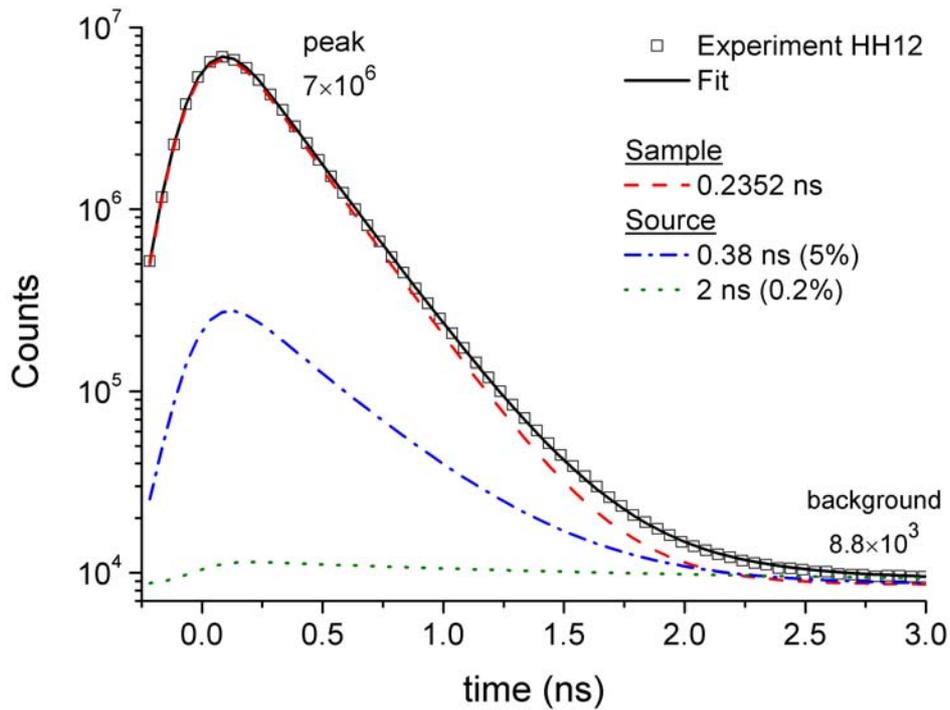

FIG. 5. Fit for PALS measurement HH12 on alloy H. The measurement was carried out at −50°C and the delay at room temperature after quenching was ≈2 min. In total 455 spectra, each accumulating for 12 minutes and yielding in total $7\times10^7$ counts were used (analysis of individual spectra, see FIG. 9). The fixed source corrections employed were ≈0.2% 2 ns and 5% 0.380 ns. The resolution function was a single Gaussian with 0.259 ns FWHM. Allowing for an additional bulk annihilation component of 0.170 ns leads to a fit containing just 1.5% of that contribution. Allowing for two free sample lifetimes produces two lifetimes that are close together. In both cases the variance of the fit is not significantly lowered by using two lifetimes for the sample.

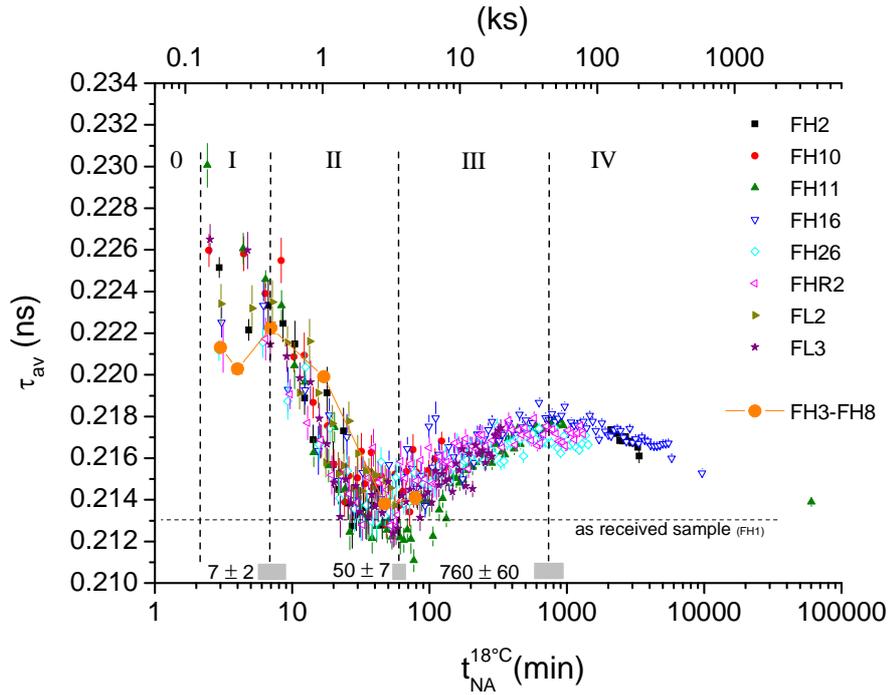

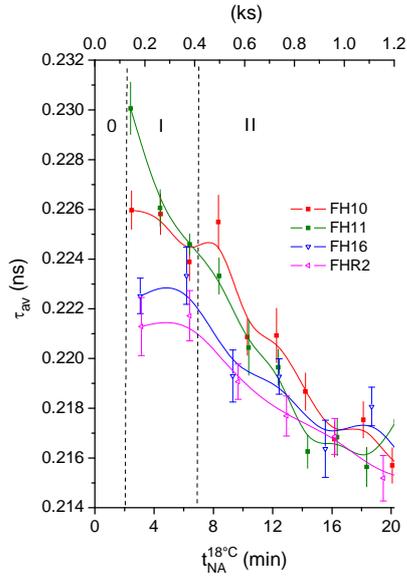

FIG. 6. Average positron lifetime for alloy group F during NA after solutionising and quenching. Measurements are from *in-situ* experiments with the exception of those shown by large filled circles that were measured *ex-situ* at –50°C. Closed/open symbols label the type of source correction applied, see Sec. II.B.4. (a) Full measured range, grey bars mark the transition from one of the stages I, II, III and IV to another. In stage 0 no data is available. Broken horizontal line: alloy in the state as-received. (b) close-up of first 20 minutes for selected data, shown on a linear scale.

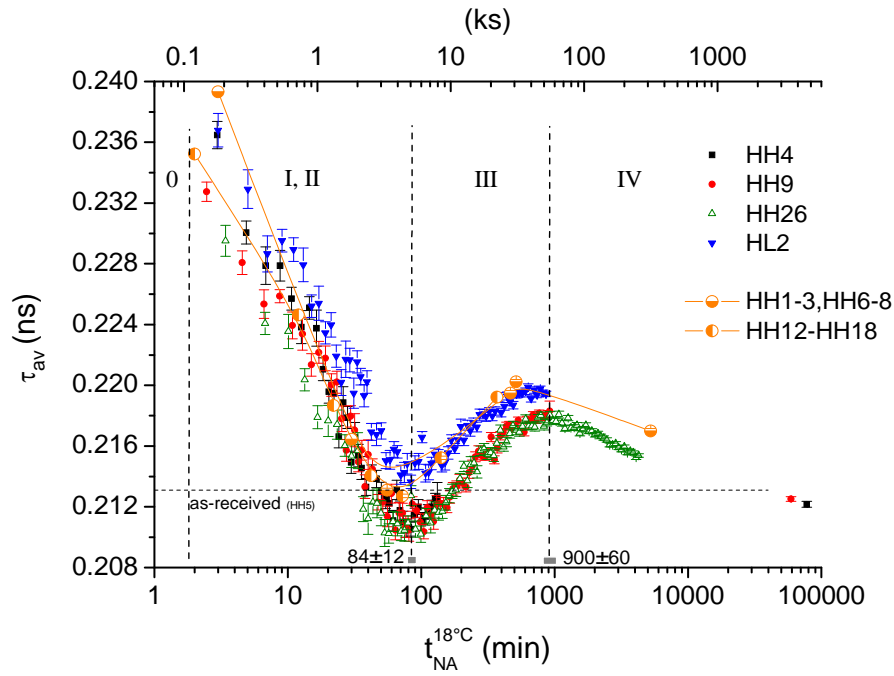

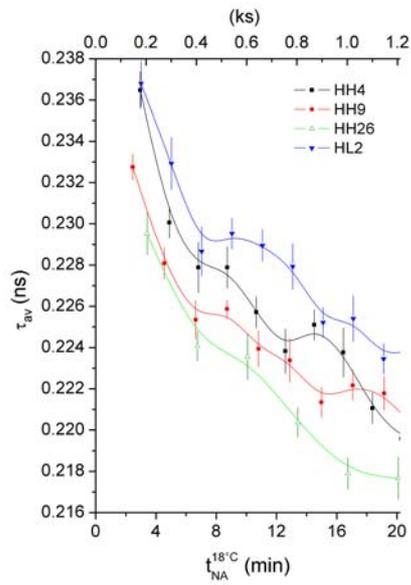

FIG. 7. Average lifetimes for alloy group H (same use of symbols as in FIG. 6).

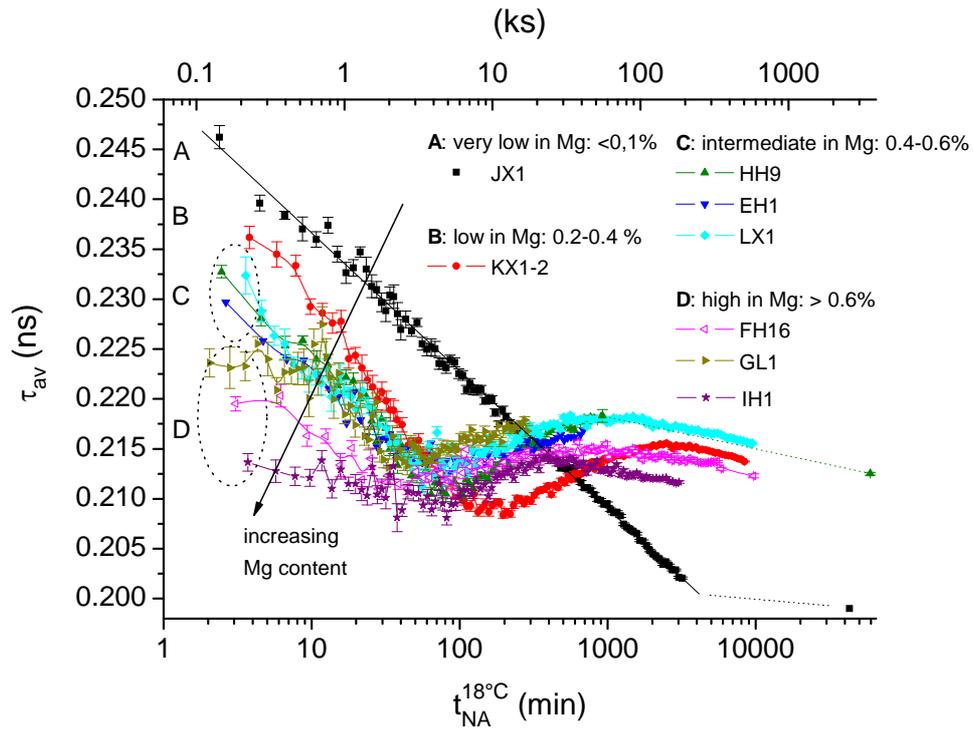

FIG. 8. Average positron lifetimes during NA at 18°C after solutionising and quenching for all the seven alloys investigated. The lines given are simple spline functions providing a guide to the eye except for alloy J where a straight line is given ($\tau_{av}= 0.250 - 0.014 \times \log(t)$ ).

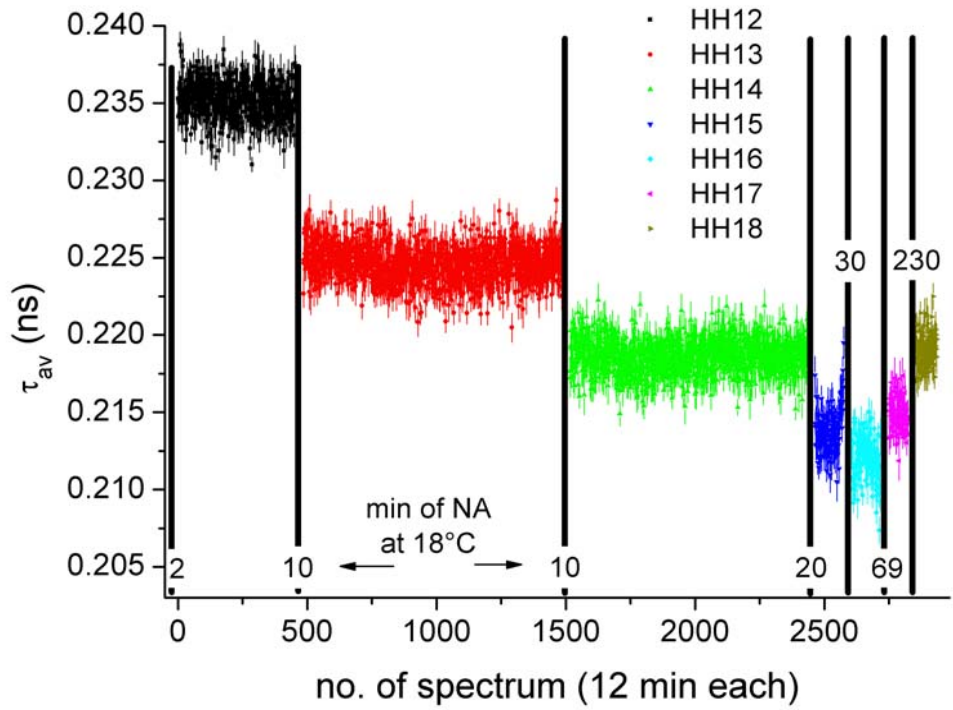

FIG. 9. Positron lifetime in alloy H measured at −50°C (measurements HH12 to HH18). The first measurement corresponds to the state frozen 2 min after solutionising and quenching. Vertical black bars denote ageing at 18°C for the times given (in min). The lifetime fit of all the accumulated data sets belonging to HH12 is shown in FIG. 5, the average of each segment in FIG. 7.

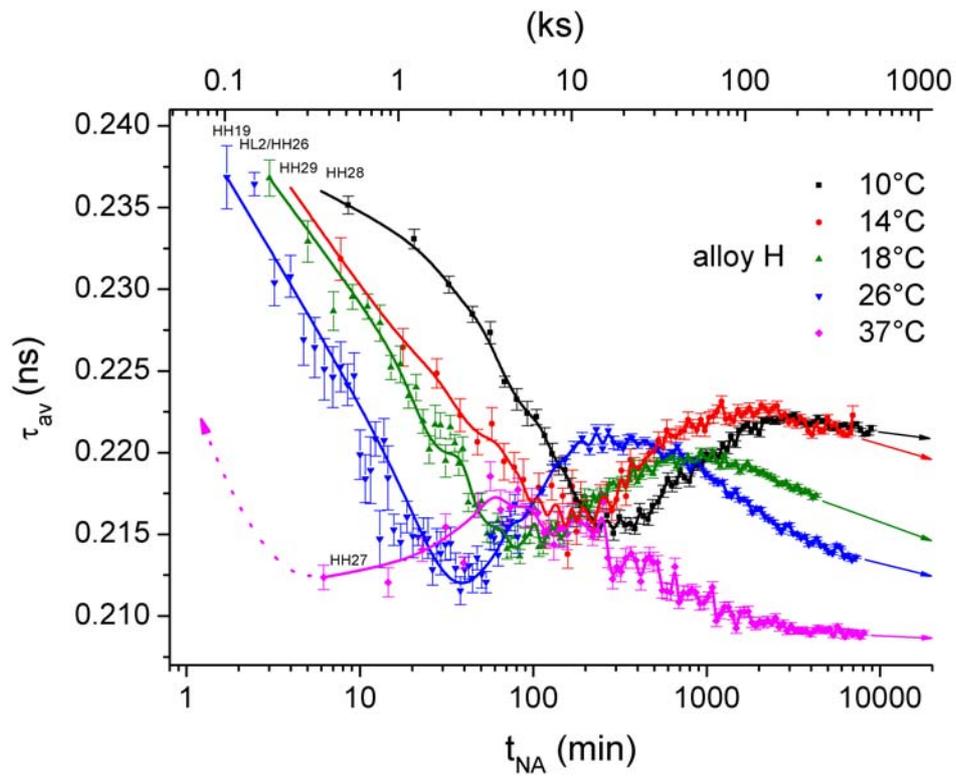

FIG. 10. Average positron life times for alloy H measured during ageing at 5 different temperatures directly after solutionising and quenching. Measured data is represented by various symbols, while lines are spline interpolations providing a guide to the eye only.

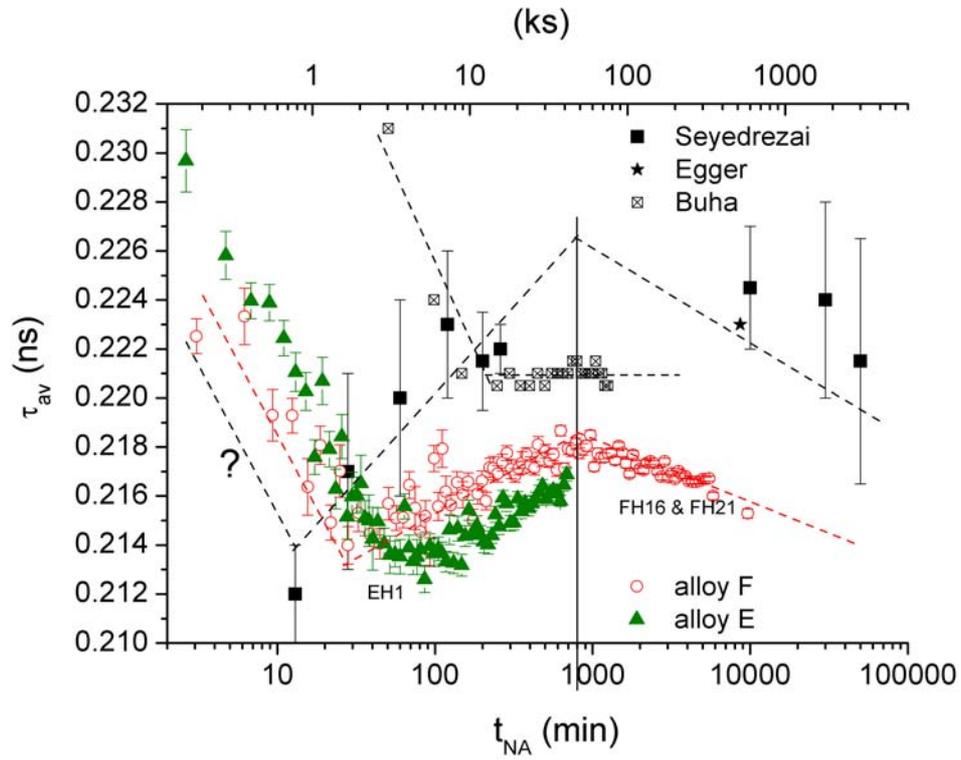

FIG. 11. Comparison of positron lifetime measurements during NA given by Seyedrezai et al. on Al-1.05wt.%Si-0.46wt.%Mg-0.14wt.%Fe alloy,[25] by Egger et al. on alloy 6082[2] and by Buha et al. on alloy 6061[26] with our measurements on the similar alloys E and F. Broken lines indicate the suspected trends of the experimental data of alloy F and that given in Refs. 25 and 26.

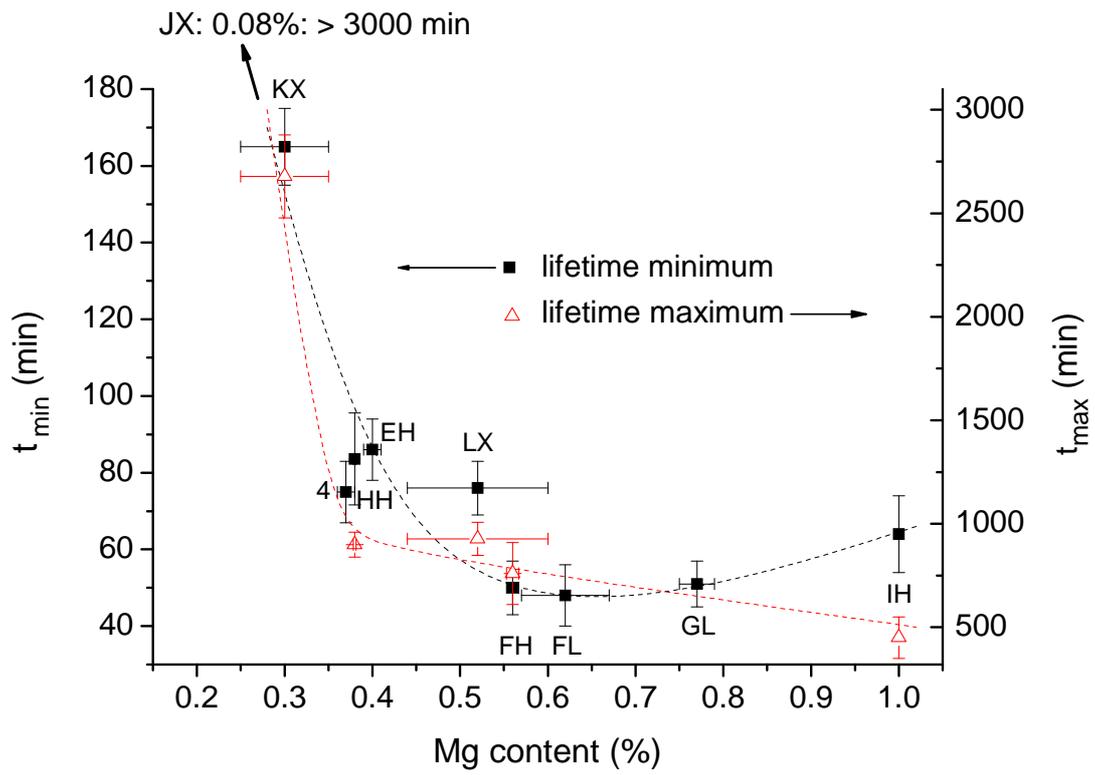

FIG. 12. Influence of the Mg content of the alloy on the time at which the minima and maxima of the positron lifetime curves occur.

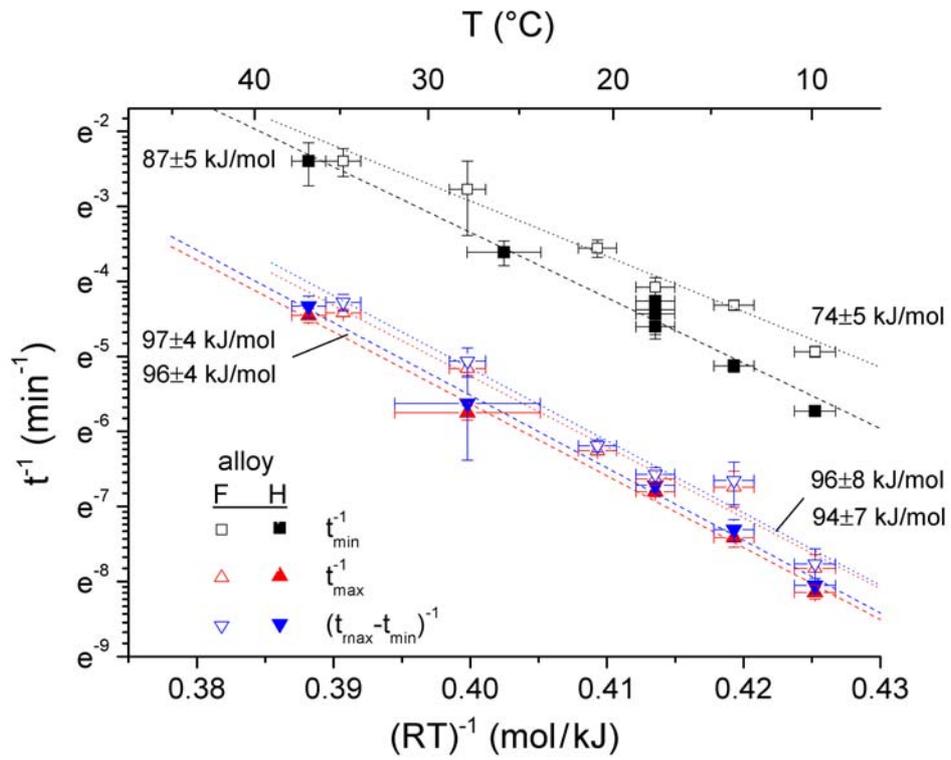

FIG. 13. Arrhenius activation energy analysis for alloy H (based on curves shown in FIG. 10) and for alloy F (based on analogous data). The 1$^{st}$ reaction is characterised by the time $t_{min}$ at which the lifetime minimum occurs. For the 2$^{nd}$ reaction, either the time $t_{max}$ of the maximum of the lifetime or the difference $t_{max}-t_{min}$ are used, see Eq.(4). The corresponding activation energies are given.

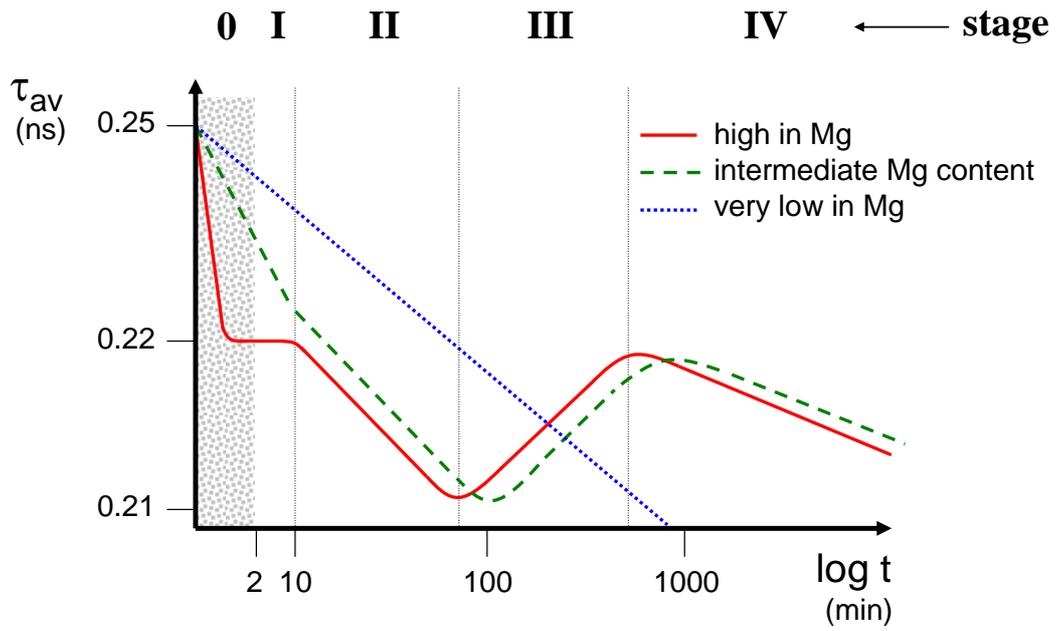

FIG. 14. Schematic representation of positron lifetime evolution at 'room temperature' after solutionising and quenching of purely ternary Al-Mg-Si alloys. Shaded area marks period in which no data are available.

**Tables**

TABLE 1. Positron lifetimes associated with different traps in a Al matrix as estimated from theoretical work and experiments found in the literature.

| type of positron trap | positron lifetime (ns) | Ref. |
|---|---|---|
| 1V in Al | $\tau_V$ = 0.240 to 0.250 | 28, 39 + other sources |
| 1V+1Mg in Al | $\tau_V$ + 0.001 to 0.003 | 35 |
| 1V+1Si in Al | $\tau_V$ – 0.001 to 0.002 | 35 |
| 1V in NN shell of Mg in Al | $\tau_V$ + 0.01 to 0.03 | 37, extrapolated and analogy to Cu assumed |
| 1V in NN shell of Si in Al | $\tau_V$ – 0.01 to 0.02 | |
| 1V + many Mg and Si in Al | unknown | |
| cluster of 60% MgSi + 40% Al in Al | $\approx$ 0.200 | 39, Eq. 6 |
| Al (bulk) | 0.160 to 170 | 28, 39 + other sources |

1V = single vacancy, 1Mg/1Si = one Mg or Si atom, NN = next neighbor

TABLE 2. Activation energies $Q_1$ and $Q_2$ for the reactions leading from stage II to III (positron lifetime minimum) and from stage III to IV (positron lifetime maximum), respectively. All values in kJ/mol.

| alloy | $Q_1$ for stage II→III | $Q_2$ for stage III→IV | |
|---|---|---|---|
| | | $t_{max}$ used | $(t_{max}-t_{min})$ used |
| H | 87 ± 5 | 96 ± 4 | 97 ± 4 |
| F | 74 ± 5 | 94 ± 7 | 97 ± 8 |